**Electronic structure of the Mott insulator $LaVO_3$ in a quantum well geometry**

Y. Hotta[a)]

*Department of Advanced Materials Science, University of Tokyo, Kashiwa, Chiba 277-8561, Japan*

H. Wadati and A. Fujimori

*Department of Physics and Department of Complexity Science and Engineering, University of Tokyo, Kashiwa, Chiba 277-8561, Japan*

T. Susaki and H. Y. Hwang[b)]

*Department of Advanced Materials Science, University of Tokyo, Kashiwa, Chiba 277-8561, Japan*

We used x-ray photoemission spectroscopy to investigate the electronic structure of the Mott insulator $LaVO_3$ embedded in $LaAlO_3$. By limiting the upper layer of $LaAlO_3$ to 3 unit cells, the underlying $LaVO_3$ could be probed. The V 2*p* core-level spectra had both $V^{3+}$ and $V^{4+}$ components, and above 2 unit cell thick $LaVO_3$, the structures exhibited spectra similar to bulk samples. The atomically flat surfaces enabled study of the emission angle dependence, which indicates the $V^{4+}$ is localized to the topmost layer. These results demonstrate the potential for probing interface electronic structure in oxide ultrathin films by surface spectroscopy.

a) Electronic mail: khotta@mail.ecc.u-tokyo.ac.jp
b) Also at: Japan Science and Technology Agency, Kawaguchi, 332-0012, Japan.

Oxide heterostructures offer the possibility to incorporate correlated electron phenomena in device geometries. In field effect devices, for example, the charge carrier density can be modulated without varying the disorder associated with substitutional chemical doping. Since the properties of correlated electron systems strongly depend on the carrier density, this approach has been used to modulate magnetic and superconducting properties by external electric field applied through the gate electrode.[1] Another example is given by rectifying oxide heterojunctions, which display a number of unusual features such as magnetically tunable junctions[2] and resistive switching.[3] In the consideration of these specific geometries, the general issue of potential electronic reconstructions of the surface/interface of a Mott insulator has been raised.[4-7]

A recent experimental approach to this problem has been *in situ* growth and photoemission spectroscopy of interfaces just below the surface, within the electron escape depth from the sample.[8] In addition to providing a powerful tool to study the interface, this technique greatly expands the range of accessible oxide materials for angle-resolved photoemission spectroscopy beyond those that cleave readily. Given the demonstrated ability to grow atomically precise heterostructures of complex oxides,[4] highly idealized interfaces can be probed. The range of study would be further extended if film growth and photoemission measurements could be separated. A chemically stable, atomically flat, and electron-transparent capping layer could be used to preserve the oxide "surface" of interest.

In this context, we have studied thin layers of the Mott insulator $LaVO_3$ embedded in $LaAlO_3$ by x-ray photoemission spectroscopy (XPS). In order to access the electronic structure of $LaVO_3$, the top layer of $LaAlO_3$ was kept at 3 unit cells (uc) in thickness. We find a systematic evolution of the vanadium core levels as the $LaVO_3$ layer is varied from 1-5 uc, and above 2 uc thickness, the core level spectra well match that of bulk samples with surfaces filed *in situ*. By measuring the emission angle dependence, we have modeled the vanadium valence profile for 3 uc thick $LaVO_3$, showing an asymmetric distribution with mixed valence $V^{4+}/V^{3+}$ in the topmost layer, and $V^{3+}$ in the lower layers.

$LaVO_3$ quantum wells with varying thickness from 1-5 uc were grown embedded in $LaAlO_3$ on atomically flat, $TiO_2$ terminated (001) $SrTiO_3$ substrates using pulsed laser deposition. $LaAlO_3$ was chosen to host the $LaVO_3$ layer since the large bandgap (5.6 eV) can effectively confine the *d*-electrons in $LaVO_3$. However, the structures were grown atop $SrTiO_3$ to more closely match bulk $LaVO_3$ (pseudocubic $LaVO_3$ = 3.93 Å, $SrTiO_3$ = 3.91 Å, pseudocubic $LaAlO_3$ = 3.78 Å). All structures were confirmed to be fully strained to the substrate by off-axis x-ray diffraction. The structures were grown at 600 °C under an oxygen partial pressure of $1 \times 10^{-6}$ Torr, with a laser fluence of 2.5 J/cm$^2$, following our previous optimization for two-dimensional layer-by-layer growth of $LaVO_3$.[9]

Figure 1(a) shows reflection high-energy electron diffraction (RHEED) oscillations during growth of the $LaAlO_3$/$LaVO_3$ heterostructures. Each oscillation corresponds to the completion of 1 uc, and the oscillating RHEED patterns indicate growth was well-controlled under the layer-by-layer mode. Figure 1(b) shows a schematic diagram of the shallow $LaVO_3$ quantum well structures grown for this study. The textured atomic force microscopy (AFM) image on the diagram, displaying the typical surface morphology of the samples, shows 1 uc steps (height ~4 Å) spaced by atomically flat terraces reflecting the miscut of the substrate.

XPS measurements were performed using a Gammadata Scienta SES-100

hemispherical analyzer and a Mg-K$\alpha$ (h$\nu$=1253.6 eV) x-ray source in a vacuum below $1.0 \times 10^{-9}$ Torr at room temperature. Figure 1(c) shows a wide scan XPS spectrum of a 50 uc LaVO$_3$ film capped with 3 uc of LaAlO$_3$. As seen by the significant V 2$p$ core-level peaks in the spectrum, the LaAlO$_3$ was thin enough to obtain the spectra from the underlying LaVO$_3$. The La 3$d$ and the Al 2$p$ core-level spectra correspond to 3+ valence states, as expected. Furthermore, the small C 1$s$ peak reflects some hydrocarbon contamination at the sample surface, caused by the exposure to atmosphere in transfer from the growth chamber to the XPS chamber.

Given the feasibility of probing LaVO$_3$ under a 3 uc LaAlO$_3$ capping layer, we investigated the systematic evolution of vanadium valence states in LaVO$_3$ quantum well structures using V 2$p$ core-level spectra for varying well thickness. Figure 2(a) shows the core-level spectra of O 1$s$ and V 2$p$, normalized to the O 1$s$ peaks for samples with the LaVO$_3$ layer varying from 1-5 uc thickness. The shoulder peaks at the higher binding energy side of the O 1$s$ peaks reflect surface hydrocarbon contamination. For the V core-level peaks, the systematic variation of the peak intensity is consistent with the thickness variation of LaVO$_3$. All of the V 2$p_{3/2}$ peaks have broad and asymmetric features toward the higher binding energy side, exhibiting the existence of not only V$^{3+}$, but also high-order valence states of vanadium. For comparison of the detailed thickness evolution of the peak shapes, the V 2$p_{3/2}$ spectra normalized to their peak areas are given in Fig. 2(b). Satellite structures of the O 1$s$ peak due to the Mg K$\alpha_5$ ghost can be seen at 512 eV, particularly for the thinnest LaVO$_3$ structures (Mg K$\alpha_{3,4}$ have been subtracted). A change in spectrum can be seen between 2 uc and 3 uc LaVO$_3$ – above 3 uc, the spectrum does not significantly evolve, and corresponds well with previous reports of bulk LaVO$_3$ filed *in situ*.[10] Indeed, the film structures appear to have more dominant $V^{3+}$ character than bulk, with fewer contributions from higher valence states (see Fig. 3).

Given the atomic flatness of our thin film structures, we can access experiments unavailable in bulk scraped samples. To probe the spatial distribution of vanadium valence states, we performed angle-dependent XPS measurements for the 3 uc $LaVO_3$ sample at emission angles $\theta$ of 30º, 55º, and 70º. The V $2p_{3/2}$ core-level spectra at these angles are shown in Fig. 3. We have fit the core-level spectra by Gaussian curves convoluted with Lorentzians. Here the full width at half maximum (FWHM) of the Lorentzian and Gaussian are 0.6 eV (Ref. 11) and 1.87 eV, respectively (the FWHM of the Gaussian arising from multiplet coupling). For each spectrum, the experimental curves were well described by two components, $V^{3+}$ and $V^{4+}$. The binding energies derived from the fits are $V^{3+}$ = 515.13 eV and $V^{4+}$ = 516.68, consistent with reported values.[12] As $\theta$ increases from 30º to 70º, the spectra are increasingly sensitive to the topmost $LaVO_3$ layer, and they increasingly exhibit a $V^{4+}$ contribution. This indicates that $V^{4+}$ is not uniformly distributed, but rather concentrated at the top of the quantum well.

To analyze the charge distribution further, Fig. 4(a) shows the fractional area ratios of the $V^{4+}$ peaks as a function of the emission angle. In the 3 uc case, the total spectral intensity for $V^{4+}$ is given by a linear combination of the photoelectron contribution from each layer,

$$I_{V^{4+}}(\theta) = \frac{a + b\exp(-d_{LVO}/\lambda\cos\theta) + c\exp(-2d_{LVO}/\lambda\cos\theta)}{1 + \exp(-d_{LVO}/\lambda\cos\theta) + \exp(-2d_{LVO}/\lambda\cos\theta)},$$

where $a$, $b$, and $c$ are the abundance ratios of $V^{4+}$ in the 1st, the 2nd, and the 3rd layer, respectively. $d_{LVO}$ is the out-of-plane lattice constant of the $LaVO_3$ film, and $\lambda$ is the mean free path of photoelectrons (∼10 Å at a photoelectron kinetic energy of ∼700 eV). We determined the best least-squares fit to the angle dependent spectra with of *a*, *b*, and *c* as free parameters. The fit indicated in Fig. 4(a) corresponds to 75% of the vanadium in the 1$^{st}$ layer as $V^{4+}$, and the remaining vanadium in the 3+ state. Beyond the 1$^{st}$ layer, the spectra are consistent with LaVO$_3$ in a $V^{3+}$ electronic state with two *d* electrons. Furthermore, the spectra of the 2 and 1 uc samples show good agreement with the tilted spectrum of the 3 uc sample at 55° and 70°, respectively.

These results indicate that high quality interfaces of thin film oxides can be preserved for surface sensitive measurements using thin epitaxial capping layers. We note that without a LaAlO$_3$ capping layer, LaVO$_3$ films showed spectra dominated by $V^{4+}$ and $V^{5+}$, as expected given the *ex situ* nature of the experiment. Furthermore, the spectra slowly evolved in time over the course of weeks and months. By contrast the capped films exhibit stable, reproducible spectra over a year after initial growth. A remaining question is the origin of the $V^{4+}$ component observed both in our capped films and in bulk. Given that all high symmetry surfaces of LaVO$_3$ are polar, this could drive reconstructions even for bulk samples with surfaces prepared *in situ*. Using the LaAlO$_3$ cap, this polar surface problem is simply translated to the LaAlO$_3$ surface, which may interact with the LaAlO$_3$/LaVO$_3$ interface. Further studies with hard x-rays with significantly larger probing depths should shed light on this issue.

We acknowledge partial support by a Grant-in-Aid for Scientific Research on Priority Areas from the Ministry of Education, Culture, Sports, Science and Technology. Y.H. acknowledges support from QPEC, Graduate School of Engineering, University of Tokyo. H.W. acknowledges support from the Japan Society for the Promotion of Science (JSPS).

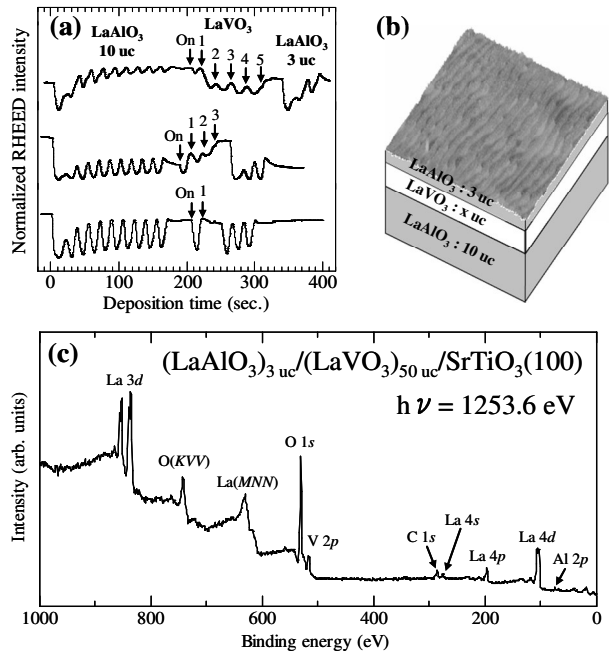

FIG. 1. (a) Normalized RHEED intensity oscillations during the growth of shallow LaVO$_3$ quantum wells of varying thickness. (b) Schematic diagram of the LaAlO$_3$/LaVO$_3$ heterostructures, and AFM image of a typical surface morphology. The step height is ~4 Å, corresponding to the height of one perovskite unit cell. (c) Wide scan XPS spectrum of the sample (LaAlO$_3$)$_{3uc}$/(LaVO$_3$)$_{50uc}$/SrTiO$_3$ (100).

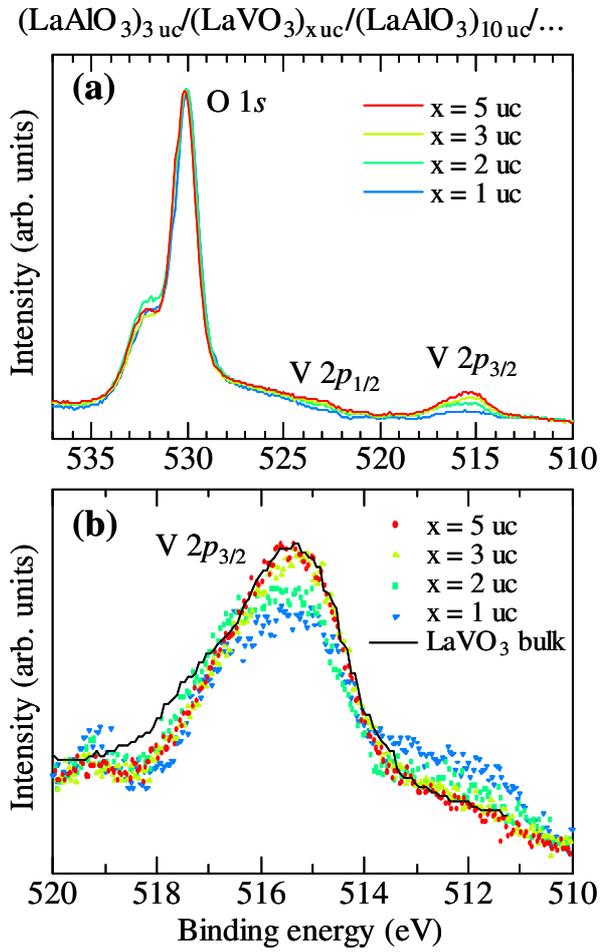

FIG. 2. (a) O 1s and V $2p_{3/2}$ photoemission spectra of $(LaAlO_3)_{3uc}/(LaVO_3)_{Xuc}/(LaAlO_3)_{10uc}$ (x = 1, 2, 3, and 5) quantum well structures, normalized to the O 1s peak height. (b) V $2p_{3/2}$ photoemission spectra normalized to the V $2p_{3/2}$ peak area. The bulk line is taken from Ref. 10, energy aligned by the O 1s peak.

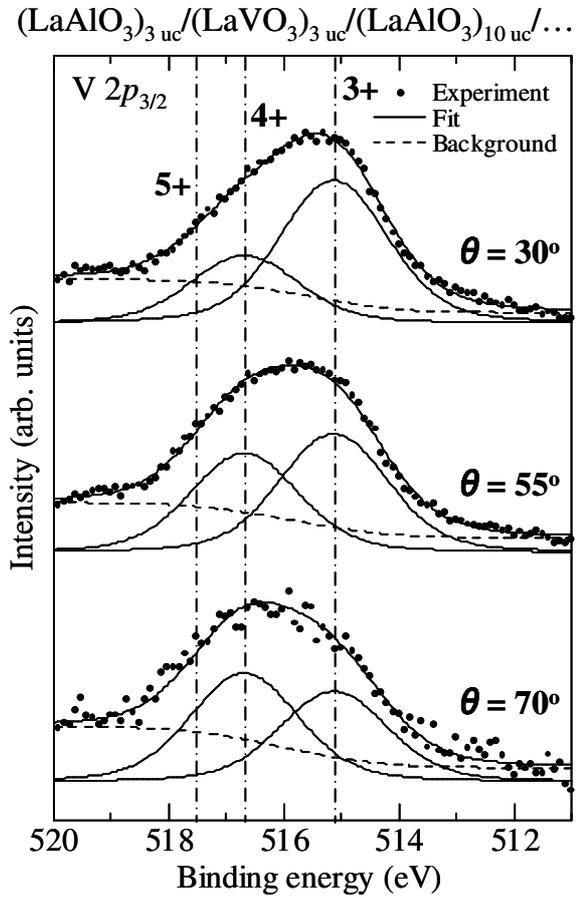

FIG. 3. Angle-dependent XPS spectra for the V $2p_{3/2}$ core-level of the $(LaAlO_3)_{3uc}/(LaVO_3)_{3uc}/(LaAlO_3)_{10uc}$ structure. Closed circles, solid lines, and dashed lines show the experimental data, the fitting curve, and the background of the spectra, respectively.

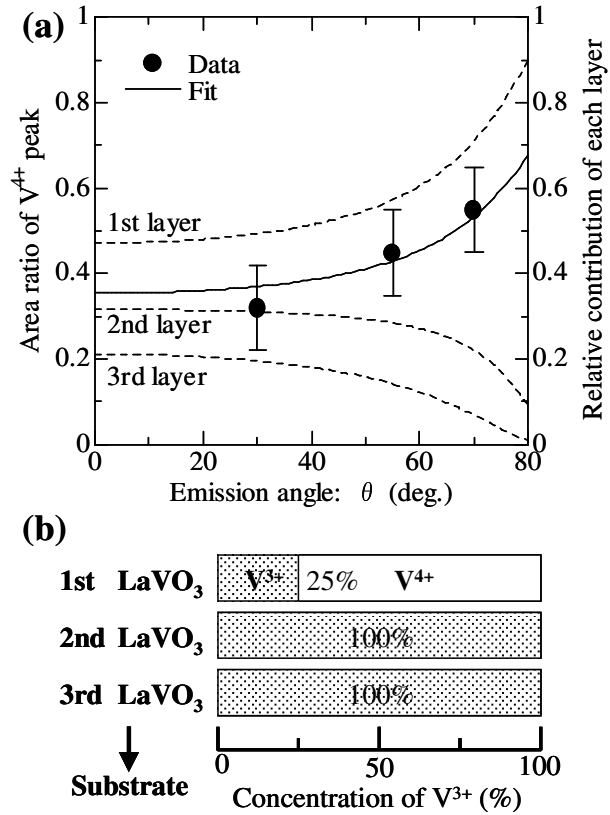

FIG. 4. (a) Fractional area ratio of the $V^{4+}$ peak to the total peak area of V $2p_{3/2}$ core-level spectra as a function of the emission angle in the $(LaAlO_3)_{3uc}/(LaVO_3)_{3uc}/(LaAlO_3)_{10uc}$ sample. Dashed lines represent the relative contribution of each layer to the total spectra. A best fit to the data corresponds to 75% $V^{4+}$ in the 1st layer, with the rest $V^{3+}$, as shown in (b).